\documentclass[nofootinbib,showpacs,preprintnumbers,amsmath,amssymb]{revtex4}
\usepackage{epsfig}

\begin{document}

\title{Synthesis of DGLAP and total resummation of leading logarithms
for the non-singlet spin structure function $g_1$}

\vspace*{0.3 cm}

author{B.I.~Ermolaev} \affiliation{Ioffe Physico-Technical
Institute, 194021
 St.Petersburg, Russia}
\author{M.~Greco}
\affiliation{Department of Physics and INFN, University Rome III,
Rome, Italy}
\author{S.I.~Troyan}
\affiliation{St.Petersburg Institute of Nuclear Physics, 188300
Gatchina, Russia}

\begin{abstract}
 The explicit expressions for the non-singlet DIS structure
function $g_1$ at small $x$ are obtained by resumming the leading
logarithmic contributions. The role played by the fits for the
initial parton densities currently used in the DGLAP on the
small-$x$ behavior of the non-singlet $g_1$ is discussed. Explicit
expressions combining DGLAP with our results are presented.
\end{abstract}


\maketitle

\section{Introduction}
The non-singlet component of the spin structure function $g_1$
have been investigated in great detail in deep
inelastic scattering (DIS) experiments. The standard theoretical
framework for studying the DIS structure functions is provided by
DGLAP\cite{dglap}. In this approach, $g_1^{NS}(x, Q^2)$ can be
represented as a convolution of the coefficient functions and the
evolved quark distributions. Combining these results with
appropriate fits for the initial quark distributions, provides a
good agreement with the available experimental data.

However, the DGLAP evolution eqs. were originally applied in a
range of large $x$ values, where higher-loop contributions to the
coefficient functions and the anomalous dimensions are small. Such
corrections are becoming essential when $x$ is decreasing, so
DGLAP should not work so well at $x\ll 1$. Nevertheless, DGLAP
predictions are in a good agreement with available experimental
data. It leads to the conclusion that the impact of the
higher-order corrections is negligibly small for the available
values of $x$. Below we use our results \cite{egt1} to show that
the impact of the high-order corrections on the $Q^2$ and $x$
-evolutions of the non-singlet structure functions is quite
sizable and bounds the region of strict applicability of DGLAP to
$x > 10^{-2}$. We also show that the reason for the success of
DGLAP at $x < 10^{-2}$ is related to the sharp $x$ -dependence
assumed for the initial parton densities, which is able to mimic
the role of high-order corrections.

The paper is organized as follows: In Sect.~2 we discuss the
difference of our approach with DGLAP. Then we compare our and the
DGLAP formulae for asymptotics of $g_1$.  In Sect.~3 we suggest a
 method to combine DGLAP with our approach in order to obtain
equally correct expressions for both large and small values of
$x$. Sect.~4 contains our conclusions.

\section{Comparison of DGLAP and our approach}
As the DGLAP -expressions for the non-singlet structure functions
are well-known, we discuss them briefly only. In this approach,
$g^{NS}_{1~DGLAP}(x, Q^2)$ can be represented as a convolution
\begin{equation}
\label{fdglap} g^{NS}_{1 DGLAP}(x, Q^2) = \int_x^1
\frac{dy}{y}C(x/y)\Delta q(y, Q^2)~
\end{equation}
of the coefficient functions $C(x)$ and the evolved quark
distributions $\Delta q(x, Q^2)$. Similarly, $ \Delta q(x, Q^2)$
can be expressed through the convolution of the splitting
functions and the initial quark densities $\delta q(x \approx 1,
Q^2 \approx \mu^2)$ where $\mu^2$ is the starting point of the
$Q^2$ -evolution. It is convenient to represent $f(x, Q^2)$ in the
integral form, using the Mellin transform:

\begin{equation}
\label{fdglapmellin} g^{NS}_{1~DGLAP}(x, Q^2) = (e^2_q/2)
\int_{-\imath \infty}^{\imath \infty} \frac{d \omega}{2\imath
\pi}(1/x)^{\omega} C(\omega) \delta q(\omega) \exp
\Big[\gamma(\omega) \int_{\mu^2}^{Q^2} \frac{d
k^2_{\perp}}{k^2_{\perp}} \alpha_s(k^2_{\perp})\Big]
\end{equation}
where $C(\omega)$ are the non-singlet coefficient functions,
$\gamma(\omega)$ the non-singlet anomalous dimensions and $\delta
q(\omega)$ the Mellin transforms of the initial non-singlet quark
densities. The standard DGLAP fits $\delta q(x)$ for the
non-singlet parton densities (see e.g. Refs.~\cite{a, v}) consist
of the terms singular when $x \to 0$ and the regular in $x$ part.
For example, the fit A of Ref.\cite{a} is chosen as follows:

\begin{eqnarray}
\label{fita}
&&\delta q(x) = N \eta x^{-\alpha}\phi(x)~, \\ \nonumber
&&\phi(x) \equiv (1 -x)^{\beta}(1 + \gamma x^{\delta})~,
\end{eqnarray}
with $N,~\eta$ being the normalization, $\alpha = 0.576$, $\beta =
2.67$, $\gamma = 34.36$ and $\delta = 0.75$. As the  term $
x^{-\alpha}$ in the rhs of Eq.~(\ref{fita}) is singular when $x
\to 0$ whereas the second one, $\phi (x)$ is regular, we will
address them as the singular and regular parts of the fit
respectively. Obviously, in the $\omega$ -space Eq.~(\ref{fita})
is a sum of the pole contributions:
\begin{equation}
\label{fitaomega} \delta q(\omega) = N \eta \Big[ (\omega -
\alpha)^{-1} + \sum_{k = 1}^{\infty} m_k \Big((\omega + k -
\alpha)^{-1} + \gamma (\omega + k +1 - \alpha)^{-1}\Big)\Big]~,
\end{equation}
with $m_k = \beta (\beta - 1)..(\beta - k + 1)/ k!$, so that the
first term in Eq.~(\ref{fitaomega}) (the leading pole) corresponds
to the singular term $x^{-\alpha}$ of Eq.~(\ref{fita}) and the
second term, i.e. the sum of the poles, corresponds to the
interference between the singular and regular terms. In contrast
to the leading pole position $\omega = \alpha$, all other poles in
Eq.~(\ref{fitaomega}) have negative values because $k - \alpha >
0$. An alternative approach was used in Refs.~\cite{ber}, by
introducing and solving infrared evolution equations with fixed
$\alpha_s$. This approach was improved in Refs.~\cite{egt1}, where
single-logarithmic contributions were also accounted for and the
QCD coupling was regarded as running in all  Feynman graphs
contributing to the non-singlet structure functions. In contrast
to the DGLAP parametrization $\alpha_s = \alpha_s(k^2_{\perp})$,
we used in Refs.~\cite{egt1} another parametrization where the
argument of $\alpha_s$ in the quark ladders is given by the
time-like virtualities of the intermediate gluons.
Refs.~\cite{egt1} suggest the following formulae for the
non-singlet structure functions:
\begin{equation}
\label{gnsint} g_1^{NS}(x, Q^2) = (e^2_q/2) \int_{-\imath
\infty}^{\imath \infty} \frac{d \omega}{2\pi\imath }(1/x)^{\omega}
C_{NS}(\omega) \delta q(\omega) \exp\big( H_{NS}(\omega) y\big)~,
\end{equation}
with $y = \ln(Q^2/ \mu^2)$ so that $\mu^2$ is the starting point
of the $Q^2$ -evolution. The new coefficient function $C_{NS}$ are
expressed in terms of new anomalous dimensions $H_{NS}$ whereas
$H_{NS}$ account for the total resummation of the double- and
single- logarithmic contributions (see Ref.~\cite{egt1} for
details).

\section{Comparison of DGLAP and our small-$x$ asymptotics}

When $x \to 0$, one can use the saddle point method in order to
estimate the integrals in Eq.~(\ref{gnsint}) and derive much
simpler expressions for the non-singlet structure functions:
\begin{equation}
\label{as} g_1^{NS} \sim e_q^2 \delta q(\omega_0) \xi^{\omega_0},
\end{equation}
with $\xi = \sqrt{Q^2/(x^2 \mu^2)}$ and with the intercept
$\omega_0 = 0.42$.  Eq.~(\ref{as}) predicts the asymptotic scaling
for the non-singlet structure functions: Asymptotically,
$g_1^{NS}$ depends on one argument $\xi$ instead of depending on
$x$ and $Q^2$ separately.

When the standard DGLAP fits, e.g. the fit of Eq.~(\ref{fita}),
are used, the asymptotics of $g^{NS}_{1~DGLAP}(x, Q^2)$ is also
the Regge-like:
\begin{equation}\label{asdglap}
g_{1~DGLAP}^{NS} \sim (e^2_q/2)
C(\alpha)(1/x)^{\alpha}\Big((\ln(Q^2/\Lambda^2))/
(\ln(\mu^2/\Lambda^2))\Big)^{\gamma(\alpha)/b}~,
\end{equation}
with $b = (33 - 2n_f)/12\pi$.

Comparison of Eq.~(\ref{as})and Eq.~(\ref{asdglap}) demonstrates
that both DGLAP and our approach lead to the Regge asymptotic
behavior in $x$. However, it is important that our intercept
$\omega_0$ is obtained by the total resummation of the leading
logarithmic contributions and without any assumption about fits
for $\delta q$ whereas the DGLAP intercept $\alpha$ in
Eq.~(\ref{asdglap}) is generated by the phenomenological factor
$x^{-0.57}$ of Eq.~(\ref{fita}) which mimics the total
resummation. In other words, the impact of the higher-loop
radiative corrections on the small-$x$ behavior of the
non-singlets is, actually, incorporated into DGLAP
phenomenologically, through the fits. It means that the singular
factors can be dropped from such fits
 when the coefficient function includes the total resummation
of the leading logarithms and therefore in this case fits for
$\delta q$ can be chosen as regular functions of $x$.

\section{Combining DGLAP with our higher-loop contributions}

 Eq.~(\ref{gnsint}) accounts for the resummation of
the double- and single logarithmic contributions to the
non-singlet anomalous dimensions and the coefficient functions
that are leading when $x$ is small. However, the method we have
used does not allow us to account for other contributions which
can be neglected for $x$ small but become quite important when $x$
is not far from 1. On the other hand, such contributions are
naturally included in DGLAP, where the non-singlet coefficient
function $C_{DGLAP}$ and anomalous dimension $\gamma_{DGLAP}$ are
known with the two-loop accuracy:

\begin{eqnarray}
\label{formdglap}
&&C_{DGLAP} = 1 + \frac{\alpha_s(Q^2)}{2\pi}C^{(1)}, \\ \nonumber
&&\gamma_{DGLAP} = \frac{\alpha_s(Q^2)}{4\pi}\gamma^{(0)} +
\Big(\frac{\alpha_s(Q^2)}{4\pi}\Big)^2 \gamma^{(1)}
\end{eqnarray}

Therefore, we can borrow from the DGLAP formulae the contributions
which are missing in Eq.~(\ref{gnsint}) by adding $C_{DGLAP}$ and
$\gamma_{DGLAP}$ to the coefficient function and anomalous
dimension of Eq.~(\ref{gnsint}). It is important to avoid a double
counting DL and SL terms common for these expressions.

In order to do so, let us consider the region of $x \sim 1$ where
the effective values of $\omega$ in
Eqs.~(\ref{fdglapmellin},\ref{gnsint}) are large. In this region
we can expand $H_{NS}$ and $C_{NS}$ into a series in $1/\omega$.
Retaining the first two terms in each series,
 we arrive at $C_{NS} = \widetilde{C}_{NS} +
O(\alpha_s^2)$, $H_{NS} = \widetilde{H}_{NS} + O(\alpha_s^3)$,
with (see Ref.~\cite{egt1} for details)

\begin{eqnarray}
\label{series}
&&\widetilde{C}_{NS} = 1 +
\frac{A(\omega)C_F}{2\pi}\big[ 1/\omega^2 + 1/2\omega\big],
\\ \nonumber
&&\widetilde{H}_{NS} = \frac{A(\omega)C_F}{4\pi} \big[ 2/\omega + 1
\big] +\Big(\frac{A(\omega)C_F}{4\pi}\Big)^2 (1/\omega)\big[
2/\omega + 1 \big]^2  + D[1/\omega + 1/2].
\end{eqnarray}
Now let us define the new coefficient functions $\hat{C}_{NS}$ and
new anomalous dimensions $\hat{C}_{NS}$ as follows:

\begin{eqnarray}
\label{newch}
&&\hat{H}_{NS} = \Big[H_{NS} - \widetilde{H}_{NS}
\Big] + \frac{A(\omega)}{4\pi}\gamma^{(0)} +
\Big(\frac{A(\omega)}{4\pi}\Big)^2 \gamma^{(1)}, \\
&&\nonumber \hat{C}_{NS} = \Big[C_{NS}^{(\pm)} - \widetilde{C}_{NS}
\Big] +
 1 + \frac{A(\omega)}{2\pi} C^{(1)}.
\end{eqnarray}

These new, "synthetic" coefficient functions and anomalous
dimensions of Eq.~(\ref{newch}) include both the total resummation
of the leading contributions and the DGLAP expressions in which
$\alpha_s(Q^2)$ is replaced by $A(\omega)$ defined in
Refs.~\cite{egt1} because the factorization of the phase space
into transverse and longitudinal spaces used in DGLAP to
parametrize $\\alpha_s$ is a good approximation for large $x$
only.


\begin{thebibliography}{99}

\bibitem{dglap}
G.~Altarelli and G.~Parisi, \emph{Nucl.~Phys.}, \textbf{B126}
(1977) 297; V.N.~Gribov and L.N.~Lipatov,
\emph{Sov.~J.~Nucl.~Phys.} textbf{15} (1972) 438; L.N.Lipatov,
\emph{Sov.~J.~Nucl.~Phys.} \textbf{20} (1972) 95;
Yu.L.~Dokshitzer, \emph{Sov.~Phys.~JETP} \textbf{46} (1977) 641.

\bibitem{egt1} B.I.~Ermolaev, M.~Greco and S.I.~Troyan,
\emph{Nucl.Phys.} \textbf{B594}B (2001)71; ibid
\textbf{B571}(2000)137; \emph{Phys.Lett.} \textbf{B579},
321,(2004); hep-ph/0503019.


\bibitem{a} G.~Altarelli, R.D.~Ball, S.~Forte and G.~Ridolfi,
\emph{Nucl.~Phys.}~\textbf{B496} (1997) 337; \emph{Acta Phys.
Polon.} \textbf{B29}(1998)1145;
\bibitem{v}A.~Vogt. hep-ph/0408244.


\end{thebibliography}
\end{document}